\documentclass[aps,prl,twocolumn,showpacs,float]{revtex4}
\usepackage{amsmath}
\usepackage{graphicx}
\usepackage{float}
\floatstyle{boxed}
\usepackage{bm}

\begin{document}

\bibliographystyle{prsty}

\title{Stability of Suspended Graphene under Casimir Force \vspace{-1mm} }
\author{E. M. Chudnovsky$^{1}$ and R. Zarzuela$^{2}$}
\affiliation{$^{1}$Department of Physics and Astronomy, Herbert H. Lehman College, The City University of New York, 250 Bedford Park Boulevard West, Bronx, NY 10468-1589, USA\\ $^{2}$Department of Physics and Astronomy, University of California, Los Angeles, California 90095, USA}
\date{\today}
\begin{abstract}
We consider graphene sheet suspended above a conducting surface. Treating graphene as an elastic membrane subjected to Casimir force, we study its stability against attachment to the conductor. There exists a critical elevation at the edges below which the central part of suspended graphene nucleates a trunk that becomes attached to the conductor. The dependence of the critical elevation on temperature and dimensions of the graphene sheet is computed. 
\end{abstract}


\maketitle

Graphene is a remarkable allotrope of carbon in the form of a honeycomb lattice \cite{graphene}. This 2D material is expected to mark a major breakthrough in the future of technology due to its unique mechanical, thermal, and electronic properties \cite{Neto,Novoselov}. Micro-electromechanical systems that involve suspended graphene should take into acount interaction of graphene with surrounding elements. One important source of such interactions is Casimir force between two conducting surfaces that orginates from quantum and thermal fluctuations of the electromagnetic field \cite{Kardar}. It has been intensively studied in recent years in application to graphene heterostructures \cite{Bordag,Drosdoff,Sernelius,Drosdoff-EPJ,Baibarac,Volokitin,Wu,Klimchitskaya}. 

\begin{figure}[ht]
\begin{center}
\includegraphics[width=8.7cm,angle=0]{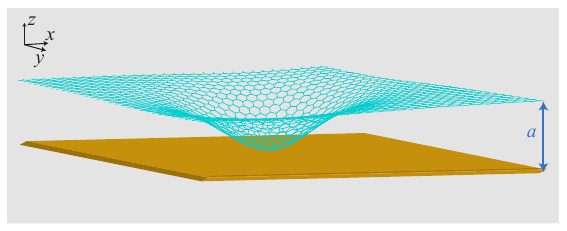}
\caption{(Color online) Suspended graphene attracted to a solid surface by Casimir force.}
\label{instability}
\end{center}
\end{figure}

In the conventional approach to Casimir interactions one studies forces between two surfaces of fixed geometry. Here we take a different approach. We treat suspended graphene as an elastic membrane and consider its deformation due to Casimir forces. Micromechanical studies  of elastic membranes have a long history \cite{Nelson-JPhys1987}. They have been recently revived in application to graphene \cite{Katsnelson,Cadelano-PRL09,Zhang-PRL11,Lindahl}. In this Letter we are concerned with stability of suspended graphene against attaching to the underlying surface. We show that there exists a critical separation from a conducting surface below which suspended graphene becomes unstable against sagging all the way down to the surface, see Fig. \ref{instability}. 

A 2D elastic membrane is described by energy \cite{Nelson-JPhys1987,Katsnelson}
\begin{equation}\label{H-membrane}
H_m = \frac{1}{2}\int d^2 r\left[\kappa ({\bm \nabla}^2h)^2 + \lambda u_{\alpha\alpha}^2 + 2 \mu u_{\alpha\beta}^2\right],
\end{equation}
where $\kappa(T)$ is the flexural stiffness constant, $\lambda$ and $\mu$ are Lam\'{e} elastic coefficients, $h({\bf r})$ is the flexural deformation perpendicular to the plane of the membrane, ${\bf u}({\bf r})$ is the displacement field in the plane of the membrane, and
\begin{equation}
u_{\alpha\beta} =\frac{1}{2}\left(\partial_{\alpha}u_{\beta} +\partial_{\beta}u_{\alpha} + \partial_{\alpha}h\partial_{\beta}h\right)
\end{equation}
is the strain tensor. The stress tensor is given by 
\begin{equation}\label{stress}
\sigma_{\alpha\beta}=\frac{\delta H_m}{\delta (\partial_{\beta}u_{\alpha})} = \lambda u_{\gamma\gamma}\delta_{\alpha\beta}+2\mu u_{\alpha\beta}.
\end{equation}

The Euler equations for ${\bf u}$ and $h$ are 
\begin{eqnarray}\label{Euler1}
&&\lambda \partial_{\alpha}u_{\beta\beta}+2\mu\partial_{\beta}u_{\alpha\beta}=0\\
\label{Euler2}
&&\kappa\nabla^{2}\nabla^{2}h-\lambda\partial_{\alpha}(u_{\beta\beta}\partial_{\alpha}h)-2\mu\partial_{\beta}(u_{\alpha\beta}\partial_{\beta}h)=0.
\end{eqnarray}
In a typical experiment a suspended graphene sheet is stretched in the $x$-direction and held by the edges running in the $y$-direction. In this case the translational invariance along the $y$-axis allows one to consider the extremal solutions of Eqs.\ (\ref{Euler1}) and (\ref{Euler2}) that depend on the coordinate $x$ only. This reduces the equations to 
\begin{eqnarray}
&& \partial_x u_{xx} =0, \quad \partial_{x} u_{yx} = 0 \\
&&  \kappa\partial_{x}^{4}h-(\lambda+2\mu)\partial_{x}(u_{xx}\partial_{x}h) = 0,
\end{eqnarray}
rendering constant values of $u_{xx}$ and $u_{yx}$ strains. The stress $\sigma$ in the $x$-direction generates the strain $u_{xx} \equiv s = \sigma/(\lambda + 2\mu)$. The equation for $h(x)$ then becomes
\begin{equation}\label{Euler-effective}
\kappa\partial_{x}^{4}h-\sigma\partial^2_{x}h = 0.
\end{equation}
It can be derived from the effective energy of the membrane 
\begin{equation}\label{effective energy}
H_{\rm eff} = \int d^2 r \left[\frac{\kappa}{2}(\nabla^{2}h)^{2}+\frac{\sigma}{2}(\nabla h)^{2}\right]
\end{equation}
that we will use below. 

Before proceeding it is useful to discuss the relative magnitude of the two terms contributing to Eq.\ (\ref{Euler-effective}) and Eq.\ (\ref{effective energy}). From their structure it is clear that the $(\nabla h)^{2}$ term dominates over the $(\nabla^2 h)^{2}$ term at curvature radii exceeding $r_c =\sqrt{\kappa/\sigma}$. The typical value of the flexural stiffness constant for graphene is of order $\kappa \sim 1$eV ($\sim 10^{-19}$J). The Lam\'{e} coefficients are in the ballpark of $10^2$J/m$^2$ and the typical elastic strain for a suspended graphene sheet is $s \sim 0.001-0.01$. This gives $\sigma \sim 0.1 - 1$J/m$^2$. Consequently, the critical curvature corresponds to $r_c \sim 1$nm. We, therefore, conclude that for all effects involving curvature radii in the excess of $1$nm the energy of a suspended graphene sheet is dominated by the second term in Eq.\ (\ref{effective energy}). 

We shall assume that the graphene sheet is suspended above a flat surface of the perfect conductor. The energy of the Casimir attraction per unit area of the graphene sheet at a distance $a$ from the conductor is \cite{Bordag}
\begin{equation}
\label{eq6}
f = -\frac{U + aV}{16 \pi a^3}
\end{equation}
with
\begin{equation}
U = \hbar c\frac{ \alpha N}{8}\left[\ln\left(1 + \frac{8}{\alpha N \pi}\right) + \frac{1}{2}\right], \quad V = k_BT\zeta(3),
\end{equation}
where $\alpha = (4\pi\epsilon_0)^{-1}e^2/(\hbar c) = 1/137.036=7.29735 \times 10^{-3}$ is the fine-structure constant, $N=4$ is the number of fermion species for graphene, and $\zeta(3) = 1.20205$ is the value of Riemann zeta function $\zeta(x)$ at $x = 3$. The crossover from the low-temperature regime with $f \propto 1/a^3$ to the high-temperature regime with $f \propto 1/a^2$ on increasing separation occurs at 
\begin{equation}
\frac{a_{t}k_BT}{\hbar c} = \frac{ \alpha N}{8\xi(3)}\left[\ln\left(1 + \frac{8}{\alpha N \pi}\right) + \frac{1}{2}\right] = 0.01508.
\end{equation}
For $T = 300$K the crossover occurs at $a_t = 115$nm, while for $T = 4$K it occurs at $a_t = 8.63\mu$m.

We will be interested in a situation when the size of the suspended graphene sheet is large compared to its distance from the underlying surface. 
\begin{figure}[ht]
\begin{center}
\includegraphics[width=9cm,angle=0]{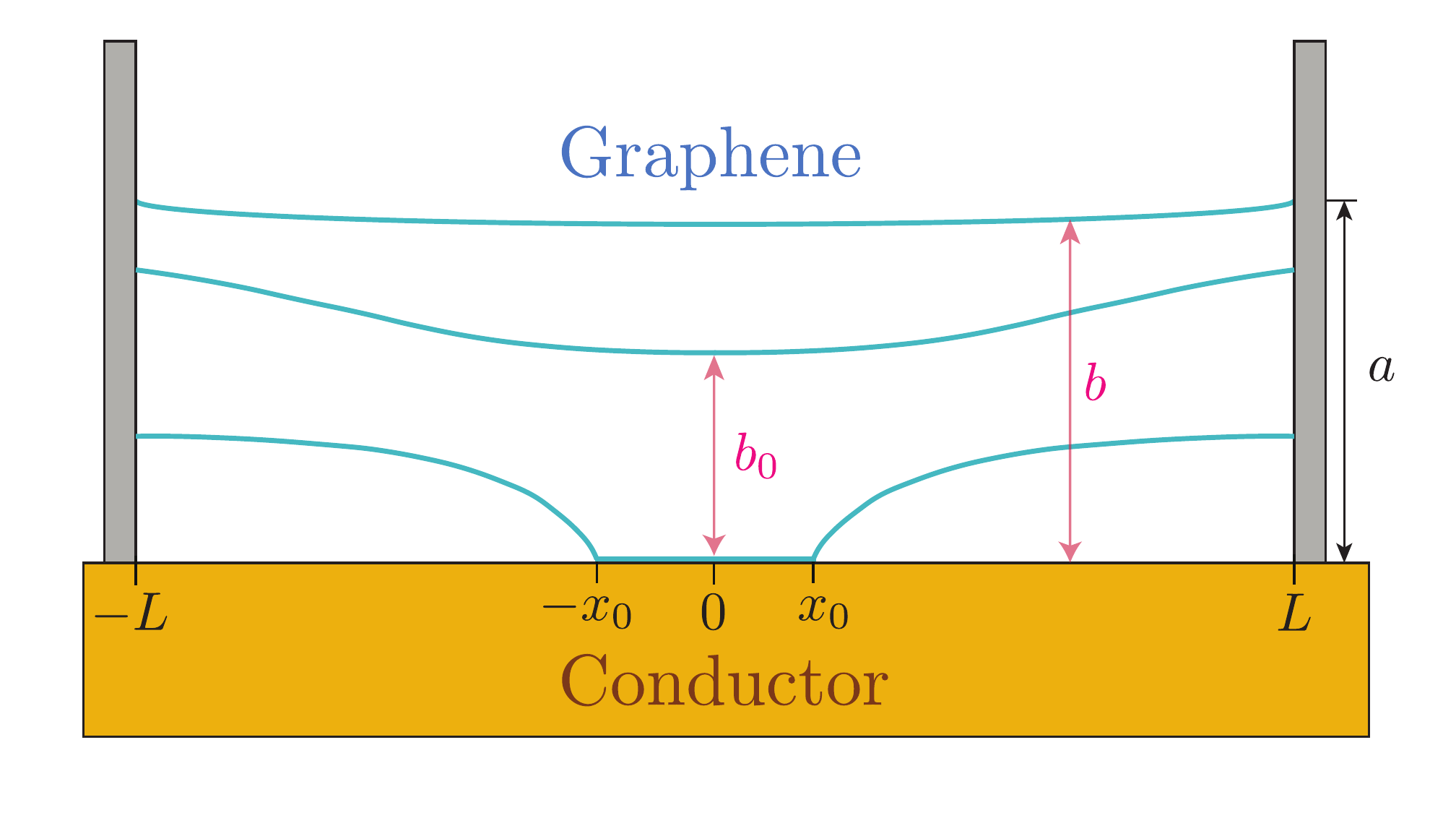}
\caption{(Color online) Casimir-driven sagging and instability of a rectangular graphene sheet against attachment to the underlying surface.}
\label{reattachment}
\end{center}
\end{figure}
The curvature of the graphene sheet will be small so that  Eq.\ (\ref{eq6}) for the energy density of Casimir attraction derived for a flat graphene must be approximately valid locally. In this case the total energy can be approximated by $H  =  H_{\rm eff} -\int d^2r f$, that is
\begin{eqnarray}\label{energy-Y}
H & = &   \frac{1}{2}\int d^2 r \left[\kappa(\nabla^{2}h)^{2}+\sigma(\nabla h)^{2}\right] \nonumber\\
&- & \beta \hbar c \int d^2 r\left[\frac{1}{(a-h)^3} + \frac{1}{a_t(a-h)^2}\right], 
\end{eqnarray}
where $a$ is a fixed distance from the conductor at the edges, $h(x,y)$ describes the sagging profile (with $h =0$ at the edges), and 
\begin{equation}\label{beta}
\beta   =  \frac{ \alpha N}{128\pi}\left[\ln\left(1 + \frac{8}{\alpha N \pi}\right) + \frac{1}{2}\right] = 0.0003606235
\end{equation}
\begin{eqnarray}\label{a_t}
a_{t}(T)  & = & \frac{16 \pi \beta}{\xi(3)} \frac{\hbar c}{k_BT}= \frac{ \alpha N}{8\xi(3)}\left[\ln\left(1 + \frac{8}{\alpha N \pi}\right) + \frac{1}{2}\right]\frac{\hbar c}{k_BT} \nonumber \\
& =& 0.01508\frac{\hbar c}{k_BT}.
\end{eqnarray}

The energy (\ref{energy-Y}) is minimized by $h({\bf r})$ satisfying the Euler equation
\begin{equation}
{\bm \nabla}^2 \frac{\delta H}{\delta {\bm \nabla}^2h} - {\bm \nabla} \frac{\delta H}{\delta {\bm \nabla}h} + \frac{\delta H}{\delta h} = 0
\end{equation}
which gives
\begin{equation}\label{Euler-full}
\left(\kappa{\bm \nabla}^4-\sigma {\bm \nabla}^2\right) h = \beta \hbar c \left[ \frac{3}{(a-h)^4} +  \frac{2}{a_t(a-h)^3}\right].
\end{equation}

Equilibrium sagging profile of a suspended graphene sheet is determined by three factors: The gain in the Casimir energy, the loss in the elastic energy, and the condition $h = 0$ at the boundary. As we shall see below, there is a critical separation at which graphene becomes unstable against attaching to the underlying surface, see Fig.\ \ref{reattachment}. 
The curvature radius of a suspended graphene must greatly exceed $r_c \sim 1$nm. This allows one to drop the first term in the left-hand side of Eq.\ (\ref{Euler-full}). In terms of dimensionless variables
\begin{equation}
\bar{\bf r} = \frac{\bf r}{a}, \quad \bar{\bm \nabla} = a{\bm \nabla}, \quad \bar{h} = \frac{h}{a}
\end{equation}
the resulting equation is
\begin{equation}
\label{Euler-strain}
{\bar{\bm \nabla}}^2 \bar{h} = -\frac{\gamma}{(1-\bar{h})^4} -\frac{\delta'_t}{(1-\bar{h})^3},
\end{equation}
where
\begin{equation}\label{factors-strain}
\gamma(a) = 3\beta\frac{ \hbar c}{\sigma a^3}, \qquad \delta_t(T) = 2\beta\frac{ \hbar c}{\sigma a^2 a_t}.
\end{equation}

\begin{figure}[ht]
\begin{center}
\vspace{0.2cm}
\includegraphics[width=8.5cm,angle=0]{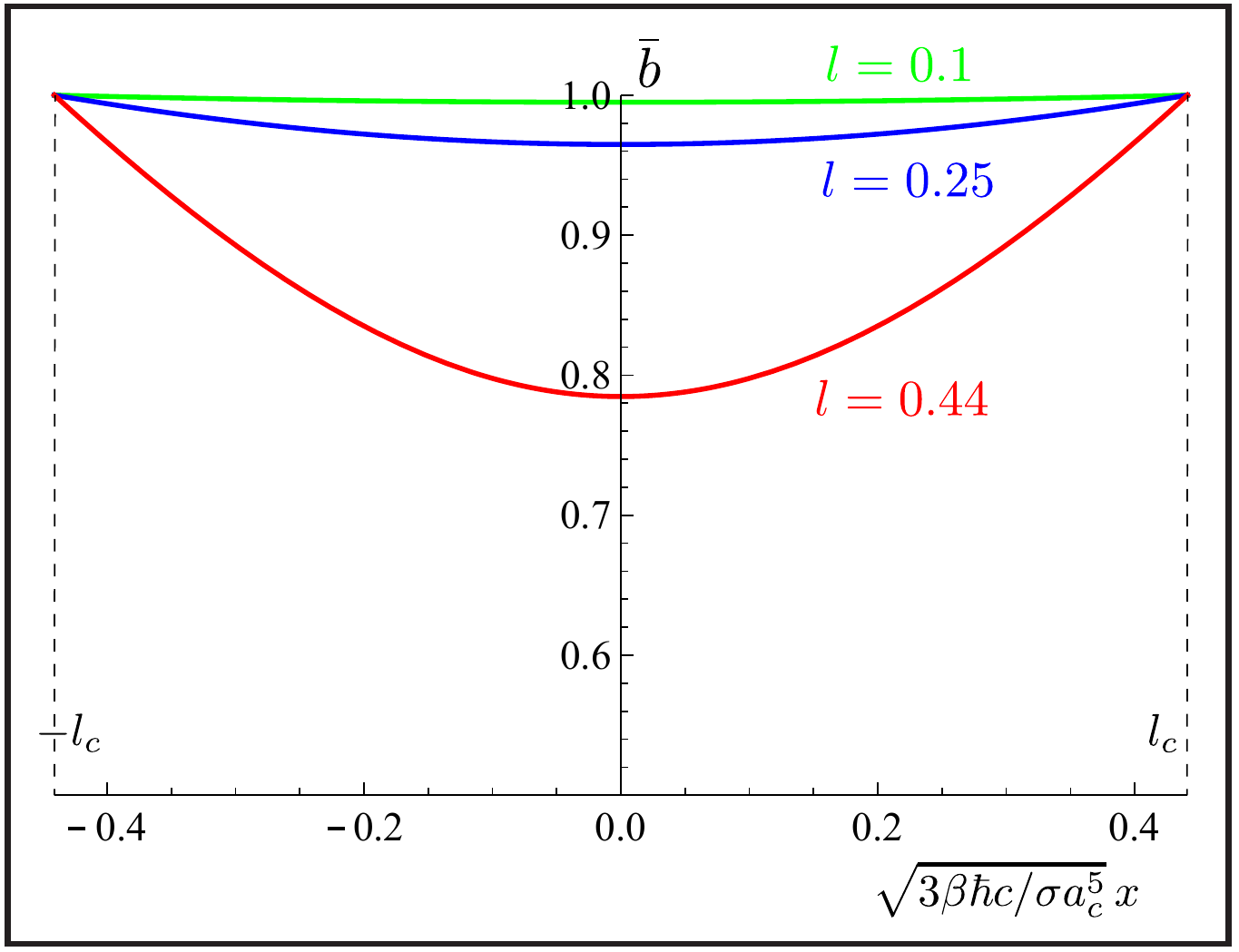}
\vspace{0.3cm}
\caption{(Color online) Profile of a sagging graphene due to Casimir force for three values of parameter $l = \gamma^{1/2}L/a$. At $l = 0.1$ the effect of Casimir attraction is weak, while at $l = 0.44$ the graphene sheet is close to the critical separation below which it becomes unstable against sagging all the way down to the conductor.}
\vspace{-0.3cm}
\label{fig-sagging}
\end{center}
\end{figure}
In the low-temperature limit, $a \ll a_t$, Eq.\ (\ref{Euler-strain}) reduces to  
\begin{equation}\label{Euler-b}
{\bar{\bm \nabla}}^2 \bar{b} = \frac{\gamma}{{\bar{b}}^4} 
\end{equation}
where $\bar{b} = 1-\bar{h}$ is the distance from the conductor in the units of $a$. We consider a rectangular graphene sheet of length $2L$, stretched in the $x$-direction by walls parallel to the $yz$-plane. In this case $h = h(x)$ and the first integral of Eq.\ (\ref{Euler-b}) is
\begin{equation}\label{eq-first}
\left(\frac{d\bar{b}}{d\bar{x}}\right)^{2}=\frac{2\gamma}{3}\left(\frac{1}{\bar{b}_{0}^{3}}-\frac{1}{\bar{b}^{3}}\right),
\end{equation}
where $\bar{b}_{0}<1$ is the minimal separation of graphene from the conductor at $\bar{x}=0$. The value of $\bar{b}_{0}$ must follow from the boundary conditions $\bar{b}(\bar{x}=\pm L/a)=1$ for the graphene sheet clipped at $x = \pm L$. 

The minimal separation at the edges, $a_c$, before the graphene sheet becomes unstable against attaching to the underlying surface (see Fig.\ \ref{reattachment}) can be estimated from the following argument. The boundary condition, $\bar{b}=1$ at $x=\pm L$, provides $\left({d\bar{b}/}{d\bar{x}}\right)^{2}\sim{(1-\bar{b}_{0})^{2}}{(a/L)^{2}}\sim ({2\gamma}/{3})\left(\bar{b}_{0}^{-3}-1\right)$,
that is, $({2\gamma}/{3})\left({L}/{a}\right)^{2}\sim \bar{b}_{0}^{3} (1-\bar{b}_{0})/({1+\bar{b}_{0}+\bar{b}_{0}^{2}})<\bar{b}_{0}<1$, which gives  $a > a_{c}\sim 2^{1/5}\left(\beta\hbar c/{\sigma}\right)^{1/5}L^{2/5}$. 

This qualitative analysis is confirmed by numerical solution of Eq.\ (\ref{eq-first}) illustrated in Fig.\ \ref{fig-sagging}. It shows that the strength of the Casimir effect is determined by the parameter $l = \gamma^{1/2}L/a = \sqrt{(3\beta \hbar c)/(a^5\sigma)} L$. The sagging profile shown in Fig.\ \ref{fig-sagging} exists at $l < l_c = 0.441$. At $l > l_c$ it is unstable against nucleation of a bubble in the central part of the graphene sheet that subsequently attaches to the underlying conductor. The exact numerical result for the critical separation at the edges in the low-temperature limit reads
\begin{equation}\label{a_c}
a_{c}= 1.73\left(\frac{\beta\hbar c}{\sigma}\right)^{1/5}L^{2/5}.
\end{equation}
Note that at the critical separation one has $a_c/L \ll 1$, that is, the graphene sheet is still close to flat in real space (when unrenormalized units of length are used). This justifies the use of Eq.\ (\ref{energy-Y}) with the Casimir potential derived for a flat graphene layer. Another important observation is that as the separation at the edges $a$ approaches $a_c$ from above, the minimum critical distance $b_0$ from the center of the graphene sheet to the underlying surface remains finite, $\bar{b}_0 = b_0/a \rightarrow 0.765$. 

For, e.g., $2L\sim2\mu$m and $\sigma\sim1$J/m$^{2}$, we get from Eq.\ (\ref{a_c}) $a_{c}\sim 11$nm. For a more macroscopic graphene sheet of $2L\sim2$mm one obtains $a_{c}\sim0.17\mu$m. These values are small compared to $a_t$ at $4$K, which justifies the low-temperature approximation used to find the critical separation in Eq.\ (\ref{a_c}). 

In the high temperature-limit, $a \gg a_t$, the first integral of Eq.\ (\ref{Euler-strain}) in one dimension is
\begin{equation}\label{eq-T}
\left(\frac{d\bar{b}}{d\bar{x}}\right)^{2}=\delta_t\left(\frac{1}{\bar{b}_{0}^{2}}-\frac{1}{\bar{b}^{2}}\right).
\end{equation}
The solution is 
\begin{equation}
\bar{b}(\bar{x})=\bar{b}_{0}\sqrt{1+{\delta_{t}\bar{x}^{2}}/{\bar{b}_{0}^{4}}}.
\end{equation}
The boundary conditions $\bar{b}(\pm L/a)=1$ give the following expression for $\bar{b}_{0}$
\begin{equation}
\label{2ndB0}
\bar{b}_{0}^{2}=\frac{1}{2}\left[1+\sqrt{1-4\delta_{t}(L/a)^{2}}\right].
\end{equation}
The stability of the solution requires $4\delta'_{t}(L/a)^{2} < 1$, which translates into 
\begin{equation}\label{ac-T}
a > a_{c}=\left[\frac{\zeta(3)k_{\textrm{B}}T}{2\pi\sigma}\right]^{1/4}L^{1/2}.
\end{equation}
As $a$ approaches $a_c$ from above, the distance from the graphene sheet to the underlying surface at the center approaches $b_0 = a/\sqrt{2} \approx 0.707a$, which is comparable to $b_0 \approx 0.765a$ in the low-$T$ limit. At $a < a_c$ graphene is unstable against its central part developing a trunk and attaching to the underlying surface, see Fig.\ \ref{reattachment}.  

Choosing $2L\sim2\mu$m, $\sigma\sim1$ J/m$^{2}$ one obtains from Eq.\ (\ref{ac-T}) $a_{c}\sim 5$nm for the critical separation at  $T=300$K. In this case $a_c \ll a_t$ and the high-temperature approximation is not justified. At $2L \gg 2$mm Eq.\ (\ref{ac-T})  gives $a_{c} \sim 0.17\mu$m $> a_t = 0.11\mu$m at $T = 300$K. Notice the week dependence of $a_c$ on temperature. This explains why the values of the critical separation in the low and high temperature regimes are similar. 

Although one can question whether our model provides a reliable approximation near the points, $x = \pm x_0$, where graphene attaches to the surface (see Fig.\ \ref{reattachment}), it is worth mentioning that it does contain such mathematical solutions. Notice that they cannot be obtained from Eqs.\ (\ref{Euler-b}) and (\ref{eq-T}) because these equations do not permit $b = 0$. However, at small radii of curvature one should use the dominant ${\bm \nabla}^4h$ term in Eq.\ (\ref{Euler-full}), that comes from the first term in Eq.\ (\ref{effective energy}). In one dimension this leads to the equation
\begin{equation}
\frac{d^4{\bar{b}}}{d{\bar{x}}^4} = -\frac{\gamma'}{{\bar{b}}^4}-\frac{\delta'_t}{{\bar{b}}^3},
\end{equation}
where
\begin{equation}\label{factors-T}
\gamma'(a) = 3\beta\frac{ \hbar c}{\kappa a}, \qquad \delta'_t(T) = 2\beta\frac{ \hbar c}{\kappa a_t(T)}.
\end{equation}
In the low-temperature regime the relevant partial solution of ${d^4{\bar{b}}}/{d{\bar{x}}^4} = -\gamma'/{\bar{b}}^4$ is 
\begin{equation}\label{profile}
\bar{b}(\bar{x}) = \left(\frac{625\gamma'}{264}\right)^{1/5}({\bar{x}-\bar{x}_0})^{4/5}, \quad \bar{x}  > \bar{x}_0.
\end{equation}
In the high-temperature limit the solution of the equation ${d^4{\bar{b}}}/{d{\bar{x}}^4} = -\delta'_t/{\bar{b}}^3$ near $x = x_0$ is
\begin{equation}
\bar{b} = (2\delta'_t)^{1/4}(\bar{x}-\bar{x}_0)\ln^{1/4}\left(\frac{1}{\bar{x}-\bar{x}_0}\right), \quad \bar{x}  > \bar{x}_0.
\end{equation}
Since $\delta_t^{1/4} \propto T^{1/4}$ the temperature dependence is week. Notice also the extremely slow divergence of the first derivative, $d\bar{b}/d\bar{x} =  (2\delta_t)^{1/4}\ln^{1/4}[1/(\bar{x}-\bar{x}_0)]$ at $\bar{x} \rightarrow \bar{x}_0$, in the high-$T$ case as compared to somewhat faster but still slow divergence $\sim 1/(\bar{x}-\bar{x}_0)^{1/5}$ in the low-$T$ case. These observations may be relevant to the problem of exfoliation of graphene. The divergence of the derivative at the separation point is contrary to the perception (derived from watching separation, e.g., of a scotch tape) that the sheet exfoliates parallel to the substrate. 

In Conclusion, by treating a suspended graphene sheet as an elastic membrane we have studied its sagging profile due to Casimir attraction to the underlying conductor. Critical separation at which graphene becomes unstable against attachment to the conductor has been computed as a function of temperature and size of the graphene sheet. While our model ignores certain effects such as, e.g., modification of the Casimir force by weak bending of the graphene sheet, it provides a reasonable estimate of the critical separation and can serve as the first approximation to the stability problem. It may also provide some hints regarding exfoliation of graphene which is the primary method of its low-cost mass production \cite{Exf_mech}.

The authors acknowledge valuable discussions with H. Ochoa. RZ thanks Fundaci\'{o}n Ram\'{o}n Areces for a postdoctoral fellowship within the XXVII Convocatoria de Becas para Ampliaci\'{o}n de Estudios en el Extranjero en Ciencias de la Vida y de la Materia.

\end{document}